\documentstyle[prd,aps,epsf,eqsecnum]{revtex}
\begin{document}
\def\CC{{\Bbb C}}
\def\NN{{\Bbb N}}
\def\QQ{{\Bbb Q}}
\def\RR{{\Bbb R}}
\def\ZZ{{\Bbb Z}}
\def\cA{{\cal A}}          \def\cB{{\cal B}}          \def\cC{{\cal C}}
\def\cD{{\cal D}}          \def\cE{{\cal E}}          \def\cF{{\cal F}}
\def\cG{{\cal G}}          \def\cH{{\cal H}}          \def\cI{{\cal I}}
\def\cJ{{\cal J}}          \def\cK{{\cal K}}          \def\cL{{\cal L}} 
\def\cM{{\cal M}}          \def\cN{{\cal N}}          \def\cO{{\cal O}}
\def\cP{{\cal P}}          \def\cQ{{\cal Q}}          \def\cR{{\cal R}} 
\def\cS{{\cal S}}          \def\cT{{\cal T}}          \def\cU{{\cal U}}
\def\cV{{\cal V}}          \def\cW{{\cal W}}          \def\cX{{\cal X}}
\def\cY{{\cal Y}}          \def\cZ{{\cal Z}}

\def\arsinh{\mathop{\rm arsinh}\nolimits}
\def\arsinh{\mathop{\rm arcosh}\nolimits}
\def\arsin{\mathop{\rm arsin}\nolimits}
\newtheorem{theorem}{Theorem}
\newtheorem{prop}{Proposition}
\newtheorem{conj}{Conjecture}
\newcommand{\be}{\begin{equation}}
\newcommand{\ee}{\end{equation}}
\newcommand{\dd}{\partial}
\newcommand{\bea}{\begin{eqnarray}}
\newcommand{\eea}{\end{eqnarray}}

\title{CLASSICAL INTEGRABILITY OF TWO DIMENSIONAL NON LINEAR SIGMA MODELS \footnote{
Preprint number: LTH 543}}

\author{N. Mohammedi \footnote{Permanent address: 
Laboratoire de Math\'ematiques et Physique Th\'eorique, 
Universit\'e Fran\c{c}ois Rabelais,
Facult\'e des Sciences et Techniques,  
Parc de Grandmont, F-37200 Tours, France.}$^{,}$\footnote{E-mail: 
nouri@celfi.phys.univ-tours.fr}}

\address{\textit{
Division of Theoretical Physics, 
Department of Mathematical Sciences,\\ 
Chadwick Building, 
The University of Liverpool, \\
Liverpool, 
L69 3BX, 
England, U.K.\\ }}

\maketitle

\vspace{5mm}

\begin{abstract}
The conditions under which a general two-dimensional non-linear sigma model is
classically integrable are given. These requirements are found by demanding that
the equations of motion of the theory are expressible as a zero curvature
relation. Some new integrable two-dimensional sigma models are then presented.

\end{abstract}

\smallskip
\smallskip

\centerline{June  2002}



\section{Introduction}

As is well-known, a two-dimensional theory is classically integrable if its equations
of motion can be represented as a zero curvature relation. That is, a  
Lax pair $\left({\cal{A}}_0,{\cal{A}}_1\right)$ can be found such that the commutator
\be
\left[\dd_0+{\cal A}_0\left(\lambda\right)\,\,,\,\,
\dd_1+{\cal A}_1\left(\lambda\right)\right] =0
\ee
yields the equations of motion of the model under consideration for all values 
of the spectral parameter $\lambda$. Here, the two-dimensional coordinates are
$\left(\tau,\sigma\right)$ with $\dd_0={\dd\over \dd\tau}$ and $\dd_1={\dd\over \dd\sigma}$.
In the rest of the paper, however, we will use the complex coordinates
$\left(z=\tau +i\sigma\,,\,\bar z=\tau -i\sigma\right)$ together with
$\dd= {\dd\over\dd z}$ and $\bar\dd={\dd\over\dd\bar z}$.
\par
The classical example of this setting is the principal chiral non-linear sigma 
model in two dimensions \cite{pohlmeyer,polyakov}. 
The action of this theory is given by
\be
S=\int {\rm d} z{\rm d}\bar{z}\,\, \eta_{kl}\,e^k_i\left(X\right) e^l_j\left(X\right)
\dd X^i \bar\dd X^j
\,\,\,\,,
\label{chiral}
\ee
where $\eta_{kl}$ is an invertible invariant bilinear form of the Lie algebra ${\cal{G}}$
defined by $\left[T_i\,\,,\,\,T_j\right]=f_{ij}^kT_k$.
That is,  $\eta_{ij}f^j_{kl} +\eta_{kj}f^j_{il}=0$.
The vielbeins $e^k_i$ satisfy the 
Cartan-Maurer relation 
\be
\dd_i e^k_j -\dd_j e^k_i + f^k_{mn} e^m_i e^n_j=0\,\,\,.  
\ee
The equations of motion of the principal chiral sigma model can be cast in the form
\be
\dd \left(e^k_i\bar\dd X^i\right) + \bar\dd \left(e^k_i\dd X^i\right)=0
\,\,\,.
\ee
This expression is reached after using the properties of $\eta_{ij}$ and $e^k_i$.
\par
The Lax pair for this model is constructed in the following manner: Consider the commutator
\bea
&\left[\dd +\left({1\over 2} +\lambda \pm\sqrt{{1\over 4}+\lambda^2}\right)
\left(e^k_i\dd X^i\right)T_k\,\,,\,\,
\bar\dd +\left({1\over 2} -\lambda \pm\sqrt{{1\over 4}+\lambda^2}\right) 
\left(e^l_j\bar \dd X^j\right)T_l\right]&
\nonumber\\
&=\left\{-\lambda
\left[\dd \left(e^k_i\bar\dd X^i\right) + \bar\dd \left(e^k_i\dd X^i\right)\right]
+\left({1\over 2} \pm\sqrt{{1\over 4}+\lambda^2}\right)
\left[\dd\left(e^k_i\bar\dd X^i\right)
-\bar\dd\left(e^k_i\dd X^i\right) +
f_{mn}^k\left(e^m_i\dd X^i\right)\left(e^n_j\bar\dd X^j\right)\right]
\right\}T_k
\,\,\,.&
\label{chirallax}
\eea
If this commutator vanishes for all values of $\lambda$ then the term proportional to
$-\lambda$ yields the equations of motion of the principal chiral sigma model.
The second term is identically zero due to the Cartan-Maurer relations between the vielbeins
$e^k_i$.
\par
In summary, the equations of motion are completely specified in terms of the currents
\be
I^j=e^j_i\dd X^i\,\,\,\,,\,\,\,\,\bar I^j_i=e^j_i \bar\dd X^i \,\,\,.
\ee
These currents satisfy the Bianchi identities $\dd \bar I^i -\bar \dd I^i +f^i_{jk}I^j\bar I^k=0$.
A convenient way of interpreting the Bianchi identities stems from the expressions
\be
\dd X^i=\left(e^{-1}\right)^i_j I^j\,\,\,\,,\,\,\,\,
\bar\dd X^i=\left(e^{-1}\right)^i_j \bar I^j \,\,\,\,,
\ee
The Bianchi identities are nothing else than the integrability conditions of these 
last relations. Namely,
\be
\dd \bar \dd X^i - \bar\dd \dd X^i=0\,\,\,.
\ee
This remark will be of use in what follows. Finally, a curvature which can be written as 
a linear combination of the equations of motion and the Bianchi identities is found.

\section{Generalisation}

We would like now to generalise the previous discussion to
any two-dimensional non-linear sigma model. Our aim is to determine 
the conditions under which a theory is integrable. 
We start from the action
\be
S=\int {\rm d} z{\rm d}\bar{z}\,\, Q_{ij}\left(X\right)\dd X^i \bar\dd X^j
\,\,\,\,.
\ee
The tensor $Q_{ij}$ has symmetric and anti-symmetric parts.
The equations of motion resulting from the variation of this action are given by
\be
{\cal{E}}_l\equiv\dd _lQ_{ij}\dd X^i \bar\dd X^j - \dd\left(Q_{lj}\bar\dd X^j\right)
- \bar\dd\left(Q_{il}\dd X^i\right)=0\,\,\,.
\ee
By contacting the equations of motion with the matrix $K^{ml}_n\left(X\right)$, one gets
\be
K^{ml}_n{\cal{E}}_l=
\left(K^{ml}_n\dd_lQ_{ij} + Q_{lj}\dd_iK^{ml}_n +Q_{il}\dd_j K^{ml}_n\right)
\dd X^i\bar\dd X^j - \dd\left(K^{ml}_nQ_{lj}\bar\dd X^j\right)
- \bar\dd\left(K^{ml}_nQ_{il}\dd X^i\right)=0\,\,\,.
\ee
In general, $K^{ml}_n$ could also
depend on some parameters which will be interpreted as the spectral parameters.
\par
Our next step is the introduction of the two currents defined by
\be
\dd X^i =\alpha^i_j\left(X\right) A^j\,\,\,\,\,\,,\,\,\,\,\,\,
\bar\dd X^i =\beta^i_j\left(X\right) \bar A^j\,\,\,\,,
\ee
where $\alpha^i_j$ and $\beta^i_j$ are two invertible matrices.
In terms of these currents, the equations of motion take the form
\be
K^{ml}_n{\cal{E}}_l=
\left(K^{ml}_n\dd_lQ_{ij} + Q_{lj}\dd_iK^{ml}_n +Q_{il}\dd_j K^{ml}_n\right)
\alpha^i_p\beta^j_q A^p\bar A^q
- \dd\left(K^{ml}_nQ_{lj} \beta^j_p \bar A^p\right)
- \bar\dd\left(K^{ml}_nQ_{il} \alpha^i_q A^q\right)=0\,\,\,.
\ee
There are various ways of expressing the equations of motion. However, this last
form is the most convenient for our purpose.
\par
The above definition of the two currents $A^i$ and $\bar A^i$ leads to some Bianchi identities.
These are found from the integrability condition 
\be
{\cal{B}}^i\equiv
\dd\bar\dd X^i-\bar\dd\dd X^i=
\dd\left(\beta^i_q \bar A^q\right)-\bar\dd\left(\alpha^i_p A^p\right)=0
\,\,\,\,.
\ee
By contracting this last equation with the matrix $L^m_{ni}\left(X\right)$, one gets
\be
L^m_{ni}{\cal{B}}^i =
 \dd\left(L^m_{ni}\beta^i_q\bar A^q\right) - \bar\dd\left(L^m_{ni}\alpha^i_pA^p\right)
-\left(\dd_jL^m_{ni} -\dd_iL^m_{nj}\right)\alpha^j_p\beta^i_qA^p\bar A^q=0
\,\,\,.
\ee
Like $K^{ml}_n$, the tensor $L^m_{ni}$ could in principle depend on some parameters too.
We have therefore expressed the equations of motion in terms of two currents
subject to some Bianchi identities. We now turn our attention to the construction of the
Lax pair.
\par
Let us consider the two differential operator as defined by
\be
{\cal{D}}^i_j=\delta^i_j\dd + V^i_j\,\,\,\,\,\,,\,\,\,\,\,\,\,
\bar{\cal{D}}^i_j=\delta^i_j\bar\dd + \bar V^i_j\,\,\,\,
\ee
where $V^i_j$ and $\bar V^i_j$ are two gauge connections. The curvature of these two
operators is given by
\be
F^i_k\equiv {\cal{D}}^i_j\bar{\cal{D}}^j_k-\bar{\cal{D}}^i_j{\cal{D}}^j_k
=\dd \bar V^i_k -\bar\dd V^i_k + V^i_j \bar V^j_k -\bar V^i_j V^j_k\,\,\,\,.
\ee
{}Following the case of the principal chiral sigma model, we require that
this curvature satisfies
\be
F^i_k= K^{il}_k{\cal{E}}_l + L^i_{kl}{\cal{B}}^l\,\,\,\,,
\label{lax}
\ee
\par
By matching the terms involving the two derivatives $\dd$ and 
$\bar\dd$ on both sides of equation (\ref{lax}), one is forced to 
choose the two gauge fields as 
\bea
V^i_k&=&\left(K^{il}_k Q_{jl}+L^i_{kj}\right)\alpha^j_m A^m
=\left(K^{il}_k Q_{jl}+L^i_{kj}\right)\dd X^j
\nonumber\\
\bar V^i_k&=&\left(-K^{il}_k Q_{lj}+L^i_{kj}\right)\beta^j_n\bar A^n
=\left(-K^{il}_k Q_{lj}+L^i_{kj}\right)\bar\dd X^j
\,\,\,.
\label{V}
\eea
Substituting  back $V^i_k$ and $\bar V^k_i$ in (\ref{lax}), leads to the 
following differential equation
\bea
&\left(K^{il}_k\dd_lQ_{mn} + Q_{ln}\dd_mK^{il}_k +Q_{ml}\dd_n K^{il}_k\right)
-\left(\dd_mL_{kn}^i-\dd_nL_{km}^i +L_{jm}^iL_{kn}^j-L_{jn}^iL_{km}^j\right)&
\nonumber\\
&=\left(K^{il}_j K^{jt}_k-K^{it}_j K^{jl}_k \right)Q_{ln}Q_{mt}
+\left(K^{il}_jL^j_{kn}- L^i_{jn}K^{jl}_k\right)Q_{ml}
+\left(K^{il}_jL^j_{km}- L^i_{jm}K^{jl}_k\right)Q_{ln}\,\,\,.&
\label{diff}
\eea
It is remarkable that this last equation is independent of $\alpha^i_j$ and $\beta^i_j$.
Therefore the choice of $A^i$ and $\bar A^i$ has no influence on the construction.
We conclude that the curvature $F^i_j$, corresponding to the 
gauge fields $V^i_j$ and $\bar V^i_j$ in  (\ref{V}),  can be written as a linear combination
of the equations of motion and the Bianchi identities if equation (\ref{diff}) is
fulfilled.
Furthermore,  the quantities $K^{ij}_k$ and $L_{jk}^i$ must depend on 
at least one spectral parameter $\lambda$. 
If the equation $F^i_k=0$ holds for all values of $\lambda$ then 
one obtains ${\cal{E}}_i=0$ and ${\cal{B}}^i=0$. 
Notice that none of the tensors $Q_{ij}$, $K^{ij}_k$ and $L^i_{jk}$ 
is, a priopri, known. Therefore, the class of two-dimensional non-linear sigma model 
which is classically integrable is specified by all the tensors $Q_{ij}$ satisfying (\ref{diff})
for some quantities  $K^{ij}_k$ and $L^i_{jk}$.
\par
So far there is no Lie algebra structure in our construction. This is due
to the fact that our gauge connections $V^i_j$ and $\bar V^i_j$ are not necessarily 
evaluated in a Lie algebra. Let us now choose a gauge connection that takes value in 
the Lie algebra ${\cal{H}}$ generated by $\left[H_i\,,\,H_j\right]=h^k_{ij}H_k$. Namely,
$W=W^iH_i$ and $\bar W=\bar W^i H_i$. In a similar manner, we require that 
\be
\left[\dd +W\,,\, \bar \dd +\bar W\right]=
\left(\dd \bar W^i -\bar\dd W^i +h^i_{jk}W^j\bar W^k\right)H_i
=\left(\mu^{il}{\cal{E}}_l +\rho_l^i{\cal{B}}^l\right)H_i
\label{curvature}
\,\,\,,
\ee
where $\mu^{il}$ and $\rho^i_l$ are the equivalent of $K^{il}_j$ and $L^i_{jl}$,
respectively. The contracted equations of motion, $\mu^{ml}{\cal{E}}_l$,
and the contracted Bianchi identities, $\rho^m_{i}{\cal{B}}^i$,  are
given by
\bea
\mu^{ml}{\cal{E}}_l &=&
\left(\mu^{ml}\dd_lQ_{ij} + Q_{lj}\dd_i\mu^{ml} +Q_{il}\dd_j \mu^{ml}\right)
\alpha^i_p\beta^j_q A^p\bar A^q
- \dd\left(\mu^{ml}Q_{lj} \beta^j_p \bar A^p\right)
- \bar\dd\left(\mu^{ml}Q_{il} \alpha^i_q A^q\right)=0\nonumber\\
\rho^m_{i}{\cal{B}}^i &=&
 \dd\left(\rho^m_{i}\beta^i_q\bar A^q\right) - \bar\dd\left(\rho^m_{i}\alpha^i_pA^p\right)
-\left(\dd_j\rho^m_{i} -\dd_i\rho^m_{j}\right)\alpha^j_p\beta^i_qA^p\bar A^q=0
\,\,\,.
\eea
A direct inspection of equation (\ref{curvature}) leads to the gauge connections
\bea
W^i &=&\left(\mu^{il}  Q_{jl}+\rho^i_j\right)\alpha^j_m A^m
=\left(\mu^{il}  Q_{jl}+\rho^{i}_{j}\right)\dd X^j
\nonumber\\
\bar W^i&=&\left(-\mu^{il}  Q_{lj}+\rho^{i}_{j}\right)\beta^j_n\bar A^n
=\left(-\mu^{il}  Q_{lj}+\rho^{i}_{j}\right)\bar\dd X^j
\,\,\,.
\label{W}
\eea
The differential equation we obtain in this case is given by
\bea
&\left(\mu^{il}\dd_lQ_{mn} + Q_{ln}\dd_m\mu^{il} +Q_{ml}\dd_n \mu^{il}\right)
-\left(\dd_m\rho^{i}_{n}-\dd_n\rho^{i}_{m}+h^i_{jk}\rho^j_m\rho^k_n
\right)&
\nonumber\\
&=h^i_{jk}\mu^{jl}\mu^{kt} Q_{ln}Q_{mt}
+h^i_{jk}\mu^{jl}\left(\rho^{k}_{m}Q_{ln}+\rho^{k}_{n}Q_{ml}\right)\,\,\,. &
\label{liediff}
\eea
Here also, the dependence on the spectral parameter $\lambda$ could only be in the tensors
$\mu^{ij}$ and $\rho^{i}_j$.
\par
There are some interesting geometrical structures appearing in equation (\ref{diff})
(and similarly in equation (\ref{liediff})). These structures are the Lie derivative
of $Q_{ij}$ with respect to $K^{il}_k$ and the curvature of the differential 
operator $\left(\nabla_i\right)^j_k= \delta^j_k \dd_i + L^j_{ki}$. Using the properties 
of these two geometrical objects, one can derive some consistency conditions for the
validity of our differential equations. According to our investigation, 
the form of these conditions is too involved and does not lead to any new insight.
Finally, it is worth mentioning that equation (\ref{liediff}) is a generalisation
of an equation that appears in the context of Poisson-Lie duality in two-dimensional
sigma models \cite{klimcik1,klimcik2}. 
This suggests a possible link between classical integrability
an T-duality in sigma models. This relashionship has been noticed 
and exploited in \cite{balog1,evans,balog2} for a very particular class of classically integrable 
sigma models.

\section{Examples}

It is easier to consider the case described by the gauge connections $W^i$ and $\bar W^i$
in (\ref{W}) 
and the differential equation (\ref{liediff}). 
Let us first check that one obtains some known integrable two dimensional non-linear
sigma models. Indeed, the Lax pair construction (\ref{chirallax})
for the principal chiral sigma model is 
found by taking
\bea
\mu^{il}&=&\lambda \eta^{ik}\left(e^{-1}\right)^l_k\nonumber\\
\rho^i_n&=&\left({1\over 2}\pm\sqrt{{1\over 4}+\lambda^2}\right)e^i_n\nonumber\\
h^i_{jk}&=&f^i_{jk}
\,\,\,\,\eea
where $\eta^{ij}$ is the inverse of the bilinear form $\eta_{ij}$.
Similarly, all the Lax pairs corresponding to the
other known integrable sigma models can be constructed in a similar manner.  
\par
By examining equation (\ref{liediff}) one can realise that there are new
integrable sigma models. 
A straight forward solution to this equation is found by taking
\bea
&& \mu^{ij}=\omega^{ik}v^j_k\nonumber\\
&& v^l_k\dd_lQ_{mn} + Q_{ln}\dd_m v^l_k +Q_{ml}\dd_n v^l_k=0\nonumber\\
&& h^i_{jk}=0\nonumber\\
&& \rho^i_j=\dd_j \Lambda^i\,\,\,\,,
\eea
where $\Lambda^i\left(X\right)$ is an arbitrary function and $\omega^{ik}$ is a constant tensor. 
This means that 
the matrix $v^l_k$ is a Killing vector of the tensor $Q_{ij}$ and the Lie algebra 
${\cal{H}}$ is Abelian.
In this case 
equations (\ref{liediff}) is satisfied for any arbitrary 
tensors $\omega^{il}$ and $\Lambda^i$. 
Notice that neither the equations of motion nor the Bianchi identities
have, for this class of sigma models, terms quadratic in the currents
$A^i$ and $\bar A^i$. 
The Lax pair construction for these types of sigma models is given by
\bea
&\left[\dd +\left(\omega^{ik} v^l_k  Q_{jl}\dd X^j+\dd \Lambda^i\right)H_i\,\,,\,\,
\bar\dd +\left(-\omega^{mn} v^r_n  Q_{rs}\bar\dd X^s+\bar\dd \Lambda^m\right)H_m\right]&
\nonumber\\
&=\left\{\omega^{mn}\left[-\dd\left(v^r_n  Q_{rs}\bar\dd X^s\right)-
\bar\dd\left(v^l_n  Q_{jl}\dd X^j\right)\right]
+\left(\dd\bar\dd \Lambda^m-\bar \dd\dd \Lambda^m\right)\right\}H_m \,\,\,.&
\eea
When this curvature vanishes, the term proportional to the arbitrary tensor $\omega^{mn}$
yields the equations of motion while the second term is identically zero. 
We have chosen here to use directly $\dd X^i$ and $\bar \dd X^i$ instead of the
currents $A^i$ and $\bar A^i$.
\par
We conclude that any two-dimensional non-linear sigma model possessing
isometries is classically integrable\footnote{Since the zero curvature condition
leads to $v^i_j{\cal{E}}_i=0$ (and not ${\cal{E}}_i=0$), we require 
the Killing vectors $v^i_j$ to be invertible.}.
In this case, for a given sigma model,
one can find an infinite number of Lax pairs: one for each choice of
$\left(\Lambda^i,\omega^{ij}\right)$. Moreover, one has a Lax pair construction involving 
an arbitrary number of spectral parameters. 
\par
An example of this class is given by a generalisation of the principal
chiral sigma model. We consider the action
\be
S=\int {\rm d} z{\rm d}\bar{z}\,\, \Omega_{kl}\,e^k_i\left(X\right) e^l_j\left(X\right)
\dd X^i \bar\dd X^j
\,\,\,\,,
\label{genchiral}
\ee
where $\Omega_{kl}$ can be any constant matrix (not necessarily the
bilinear form $\eta_{kl}$). The underlying Lie algebra is still 
${\cal{G}}$, defined in the introductory section. 
The Killing vector $v^i_j$ corresponding to the tensor 
$Q_{ij}=\Omega_{kl}\,e^k_i e^l_j$ is given by
$v^i_j=R_j^k\left(e^{-1}\right)^i_k$, where $R^i_j$ is defined by
$g^{-1}T_ig=R^j_i T_j$ and verifies the relations
$\dd_k R^i_j=f^i_{mn}R^m_j e^n_k$ and $f^i_{jk} R^j_m R^k_n=f^j_{mn} R^i_j$.
Here $g\left(X\right)$ is an element in the Lie group
corresponding to the Lie algebra ${\cal{G}}$.
The vielbeins are given by the usual experession 
$e^i_j T_i=\left(g^{-1}\dd_j g\right)^iT_i$.
\par
At this point some remarks are due. Firstly, since 
the  principal chiral sigma model is a special case of the action (\ref{genchiral}),
we have therefore found a new Lax pair formulation for this model. 
Secondly, there is another model based on the Lie algebra ${\cal{G}}=SU\left(2\right)$
and which is known to be classically integrable \cite{cherednik}. This 
model corresponds to taking a diagonal $\Omega_{ij}$ with
$\Omega_{11}=\Omega_{22}\ne \Omega_{33}$. In this case also our Lax pair construction
is different from that of \cite{cherednik}. 
Finally, we should mention that an earlier attempt to study the classical integrability
of the model (\ref{genchiral}) did not produce any new integrable systems \cite{sochen}.
\par
The second example we consider here is described by the action
\be
S=\int {\rm d} z{\rm d}\bar{z}\,\, Q^{ij}\left(\chi\right)\dd \chi_i \bar\dd \chi_j
\,\,\,\,,
\ee
where $Q^{ij}$ is given through its inverse by
\be
Q_{ij}=\eta_{ij} +f^k_{ij}\chi_k\,\,\,.
\ee
This model is the non-Abelian T-dual of the chiral principal sigma model 
\cite{nadual1,nadual2,nadual3,nadual4}.
It is not known whether this model is classically integrable. 
The study of this model sheds, therefore,  some light
on whether duality preserves integrability.
\par
The equations of motion stemming from our action can be cast in the form
\be
\dd \bar J^l - \bar \dd J^l + f^l_{ij} J^i\bar J^j=0\,\,\,,
\ee
where the two currents are given by \cite{amit}
\be
J^l=Q^{il}\dd\chi_i\,\,\,\,\,,\,\,\,\,\,
\bar J^l=-Q^{lj}\bar\dd\chi_j\,\,\,.
\ee
This definition leads to the Bianchi identities ($\dd\bar\dd \chi_i -\bar\dd\dd\chi_i=0$)
\be
\dd \bar J^j + \bar \dd J^j + 
\eta^{jk}f_{kl}^i\chi_i\left(\dd \bar J^l - \bar \dd J^l + f^l_{mn} J^m\bar J^n\right)
=0\,\,\,.
\ee
The Lax pair construction is found through the commutator
\bea
&\left[\dd +\left({1\over 2} +\lambda \pm\sqrt{{1\over 4}+\lambda^2}\right)
J^kT_k\,\,,\,\,
\bar\dd +\left({1\over 2} -\lambda \pm\sqrt{{1\over 4}+\lambda^2}\right) 
\bar J^lT_l\right]&
\nonumber\\
&=\left\{-\lambda\left[
\dd \bar J^j + \bar \dd J^j + 
\eta^{jk}f_{kl}^i\chi_i\left(\dd \bar J^l - \bar \dd J^l + f^l_{mn} J^m\bar J^n\right)\right]
\right. & \nonumber\\
&+\left[\left({1\over 2} \pm\sqrt{{1\over 4}+\lambda^2}\right)\delta^j_l
+\lambda \eta^{jk}f_{kl}^i\chi_i\right]
\left(\dd \bar J^l - \bar \dd J^l + f^l_{mn} J^m\bar J^n\right)
\left.\right\}T_j
\,\,\,.&
\eea
If this commutators is to vanish then the term propotional
$-\lambda$ yields the Bianchi identities while the term proportional to
$\left[\left({1\over 2} \pm\sqrt{{1\over 4}+\lambda^2}\right)\delta^j_l
+\lambda \eta^{jk}f_{kl}^i\chi_i\right]$ leads to the equations of motion.
Therefore, the non-Abelian dual of the principal chiral sigma model is also
classically integrable.
\par
In summary,  we have given in this paper the necessary constraints for a non-linear sigma model to
be classicaly integrable. The equations found are not empty and lead to some unexpected
integrable theories. The geometrical nature of these equations might be employed to
find more solutions.

\smallskip
\smallskip
\smallskip
\bigskip

\noindent $\underline{\hbox{\bf Acknowledgments}}$: I would like to thank the members of the 
Division of Theoretical Physics at the University of Liverpool for their hospitality. 
I would like also to thank P\'eter Forg\'acs, Ian Jack , Tim Jones and Max Niedermaier for very
useful discussions.
This research is supported by the CNRS.

\end{document}